# Aggregate Hazes in Exoplanet Atmospheres

Danica Adams[1,2,*], Peter Gao[2,3], Imke de Pater[2], and Caroline V. Morley[4]

[1]Division of Geological and Planetary Sciences, California Institute of Technology, 1200 E. California Blvd. MC 150-21, Pasadena, CA 91125, USA
[2]Astronomy Department, University of California, Berkeley, 501 Campbell Hall, MC 3411, Berkeley, CA 94720-3411, USA
[3]51 Pegasi b Fellow
[4]Department of Astronomy, University of Texas at Austin, 2515 Speedway, Austin, TX 78712

*Corresponding author. Email: djadams@caltech.edu

## Abstract

Photochemical hazes have been frequently used to interpret exoplanet transmission spectra that show an upward slope towards shorter wavelengths and weak molecular features. While previous studies have only considered spherical haze particles, photochemical hazes composed of hydrocarbon aggregate particles are common throughout the solar system. We use an aerosol microphysics model to investigate the effect of aggregate photochemical haze particles on transmission spectra of warm exoplanets. We find that the wavelength dependence of the optical depth of aggregate particle hazes is flatter than for spheres since aggregates grow to larger radii. As a result, while spherical haze opacity displays a scattering slope towards shorter wavelengths, aggregate haze opacity can be gray in the optical and NIR, similar to those assumed for condensate cloud decks. We further find that haze opacity increases with increasing production rate, decreasing eddy diffusivity, and increasing monomer size, though the magnitude of the latter effect is dependent on production rate and the atmospheric pressure levels probed. We generate synthetic exoplanet transmission spectra to investigate the effect of these hazes on spectral features. For high haze opacity cases, aggregate hazes lead to flat, nearly featureless spectra, while spherical hazes produce sloped spectra with clear spectral features at long wavelengths. Finally, we generate synthetic transmission spectra of GJ 1214b for aggregate and spherical hazes and compare them to space-based observations. We find that aggregate hazes can reproduce the data significantly better than spherical hazes, assuming a production rate limited by delivery of methane to the upper atmosphere.

## 1 Introduction

Exoplanet transmission spectra that display subdued molecular features and scattering slopes are seen across a wide swath of planet types, from Super Earths to hot Jupiters (e.g., Sing et al., 2016; Crossfield and Kreidberg, 2017). Aerosols have been used to interpret these observations, but their origins and composition are uncertain. In general, it has been suggested that hazes composed of small (<~0.1 microns) spherical particles, possibly stemming from photochemistry, produce the slopes in the optical and NIR regions of transmission spectra, while condensation clouds with larger spherical particles (~1 micron) flatten spectra across all wavelengths (e.g. Barstow et al. 2017).

Photochemical hazes have been observed on many solar system bodies, including sulfuric acid aerosols in the upper haze of Venus, stratospheric sulfate aerosols on Earth (e.g., Lazrus and Gandrud, 1974; Turco et al., 1982), and hydrocarbon hazes on the giant planets (e.g., Wong et al., 2003; Koskinen et al., 2016), Titan (e.g., Rannou et al., 2010), Pluto (e.g., Gladstone et al., 2016), and Triton (e.g., Hillier et al., 1991). While some of these hazes are made up of spherical particles (e.g. Kawabata et al. 1980), others are composed of fractal aggregates - loose collections of smaller "monomers" that are highly porous and irregular in shape described by (e.g., Forrest and Witten, 1979; Sorensen and Roberts, 1997):

$$N = k_o \left(\frac{R}{r_m}\right)^{D_f}, \qquad (1)$$

where R represents the radius of the fractal aggregate that is composed of N monomers, each of a radius $r_m$. The structural coefficient, $k_o$, is a prefactor that affects the fractal scaling relationship and the description of the particles' radiative properties. We assume $k_o$ to be of order unity, as previous works have also done (e.g., West et al., 1991; Wolf et al., 2010). $D_f$ is the fractal dimension of the aggregate, which describes the porosity of the fractal. Zhang et al. (2013) investigated stratospheric aerosols on Jupiter using multiple-phase-angle images from the Cassini Imaging Science Subsystem (ISS) and ground-based near infrared (NIR) spectra. Stratospheric haze at 10-50 mbar, depending on latitude, was found to be composed of fractal aggregate particles consisting of a thousand 10-nm sized monomers. At Titan, Tomasko et al. (2008) analyzed measurements from Huygens' Descent Imager/Spectral Radiometer (DISR). The aerosols' vertical distribution, phase function, and single scattering albedo were determined by multi-directional measurements from the UV photometers, and the single-scattering phase function of the aerosols between 30 and 80 km were found to be consistent with fractal aggregates. In the lower 30 km of Titan's atmosphere, the wavelength dependence decreased compared to that at higher altitudes, suggesting that the aerosols continued to grow with decreasing altitude. Robinson et al. (2014) analyzed spectral observations at Titan made by the Cassini spacecraft, and found that the aerosol distribution reached unit optical depth at pressures less than 0.1-10 mbar, depending on wavelength, which is comparable to the pressures probed by exoplanet transmission spectra. Gao et al. (2017a) compared numerical models of both spherical and aggregate hazes to New Horizons observations of UV extinction in Pluto's atmosphere. They determined that Pluto's photochemical hazes are composed primarily of fractal aggregate particles, though spherical particles may also be present, as inferred from forward scattering observations (Cheng et al. 2017).

Since fractal aggregates are common in solar system hazes, it is possible that exoplanet hazes may be similar. Kopparla et al. (2016), for example, examined the effects of aggregate hazes on the polarization of reflected light from giant exoplanets. However, they assumed a simplified haze layer in their model. Hazes, and aerosols in general, are controlled by microphysical processes that sculpt their particle size and spatial distributions, but haze models that have considered microphysical processes in the context of exoplanets have mainly examined the effect of spherical haze particles. Lavvas and Koskinen (2017) applied a 1D aerosol microphysics model to produce aerosol distributions on HD 209458b and HD 189733b. The aerosol distribution was found to depend on the particle composition, photochemical production rate, and atmospheric mixing. Soot aerosols were found to match the primary transit observations of HD 189733b. Morley et al. (2013) calculated the vertical profile of hydrocarbons in the atmosphere of GJ 1214b to derive haze particle distributions that could match the observed flat spectra. Particle sizes of 0.01-0.25 $\mu m$ were found to best match the data, but condensation clouds were also considered. Kawashima and Ikoma (2018) developed a numerical model to simulate the production, growth, and settling of hydrocarbon haze particles on warm exoplanets (<1000 K). The size distribution of haze particles was fairly broad, and they concluded that the breadth of observed exoplanet spectra could be caused by variations in the production rate of haze monomers due to the varying UV irradiation intensity of the host stars.

In this study, we apply an aerosol microphysics model that considers transport and coagulation over a multidimensional phase space to present a comprehensive analysis of haze opacity in giant exoplanet atmospheres. Specifically, this is the first study to consider the microphysics of aggregate haze particles in exoplanet atmospheres. We also apply our model to the super-Earth GJ 1214b, which has been observed extensively using space- and ground-based observatories, (e.g. Gillon et al., 2014, Kreidberg et al., 2014, Bean et al., 2010), and many observations reported flat spectra. Kawashima and Ikoma (2018) modeled spherical haze particles in the atmosphere of GJ 1214b but required a haze forming efficiency that was orders of magnitude greater than that of Titan in order to produce a flat spectrum, while at low haze forming efficiencies a sloped spectrum at optical wavelengths was produced due to haze scattering, with spectral absorption features of molecular species visible at longer wavelengths. Since spherical hazes did not fit the observations well when considering a haze forming efficiency comparable to that in the solar system, we consider the effect of aggregate hazes on GJ 1214b's transmission spectra.

We describe our methodology in section 2. In section 3, we analyze the effect of varying haze particle shape, monomer production rate, eddy diffusion coefficient, and monomer size on particle size distributions and haze opacity. In section 4, we discuss the impact of our assumptions and the implications of our results on transmission spectra of giant exoplanets and GJ 1214b. We present our conclusions in Section 5.

## 2 Model

We calculate the equilibrium haze particle size distribution using the 1-D Community Aerosol and Radiation Model for Atmospheres (CARMA; Turco et al., 1979; Toon et al., 1988; Jacobson and Turco, 1994; Ackerman et al. 1995; Bardeen et al., 2008; Wolf and Toon, 2010). CARMA solves the continuity equation of aerosol particles that experience production via particle nucleation, growth via condensation and coagulation, loss via evaporation, and transport. We follow the methodology outlined in Gao et al. (2017a) for Pluto hazes to simulate haze distributions in exoplanet atmospheres, where nucleation, condensation, and evaporation were ignored due to the low volatility of the haze material and the uncertain chemical pathway leading to haze formation. The change with time of haze particle number density, $n_p(z)$, in the $p$th mass bin at altitude $z$ is given by:

$$\frac{\partial n_p}{\partial t} = \frac{1}{2}\sum_{i=1}^{i=p-1} K_{i,p-i} n_i n_{p-i} - n_p \sum_{i=1}^{i=N} K_{i,p} n_i - \frac{1}{z^2}\frac{\partial(z^2\Phi)}{\partial z} + \delta_{p,1}\delta_{z,z_{top}} P \qquad (2)$$

where $K_{i,j}$ is the Brownian coagulation kernel between particles in mass bins $i$ and $j$ and P is the production rate of the minimum mass particles at the top of the atmosphere, as represented by the Kronecker deltas, which specify that production only occurs at 1 microbar at the top of the atmosphere. The first term on the right hand side of Eq. (2) represents the increase in $n_p$ due to the coagulation of smaller particles with total mass equal to that of particles in the $p$th mass bin; the second term represents the decrease in $n_p$ due to coagulation of particles in the $p$th mass bin with other particles to generate more massive particles; and the third term represents vertical transport with $\Phi$ as the particle flux, defined as:

$$\Phi = -w_{sed} n - K_{zz} n \frac{\partial(n_p/n)}{\partial z}, \qquad (3)$$

with $w_{sed}$ as the sedimentation velocity, $\frac{\partial(n_p/n)}{\partial z}$ as the gradient in mixing ratio of the haze particles, and $K_{zz}$ as the eddy diffusion coefficient. Note that all variables in Eq. (3) are functions of $z$. We refer the reader to the appendix of Gao et al. (2018a) for a full description of $w_{sed}$ and $K_{i,j}$.

Each mass bin corresponds to particle masses twice that of the previous bin. We use 47 bins in our model, with the mass in the first bin corresponding to two monomers in the aggregate case and an equivalent mass in the spheres case. We assume a zero concentration lower boundary condition at 10 bars to simulate loss by thermal decomposition. The haze material mass density is taken to be 1 g/cm3, which is typical of organics including both hydrocarbon soots (e.g., Maricq et al., 2004; Rissler et al., 2013) and tholins (e.g., Trainer et al., 2006; Horst and Tolbert, 2013). The fractal dimension of aggregate particles is assumed to vary with the number of monomers per aggregate, as shown in Fig. 1, following the parametrization of Wolf and Toon (2010), where the smallest aggregates, with only two monomers, have a $D_f$ of ~1.5. Small aggregates have a low fractal dimension. This corresponds to "chain growth," in which monomers coagulate to form linear chain-like structures. As particles grow larger (~1000 monomers), they restructure and collapse into more compact arrangements, resulting in a higher fractal dimension. We set $D_f$ to a constant 2.4 for particles with more than ~2000 monomers, an extrapolation of the parameterization of Wolf and Toon (2010) to larger aggregates than they considered.

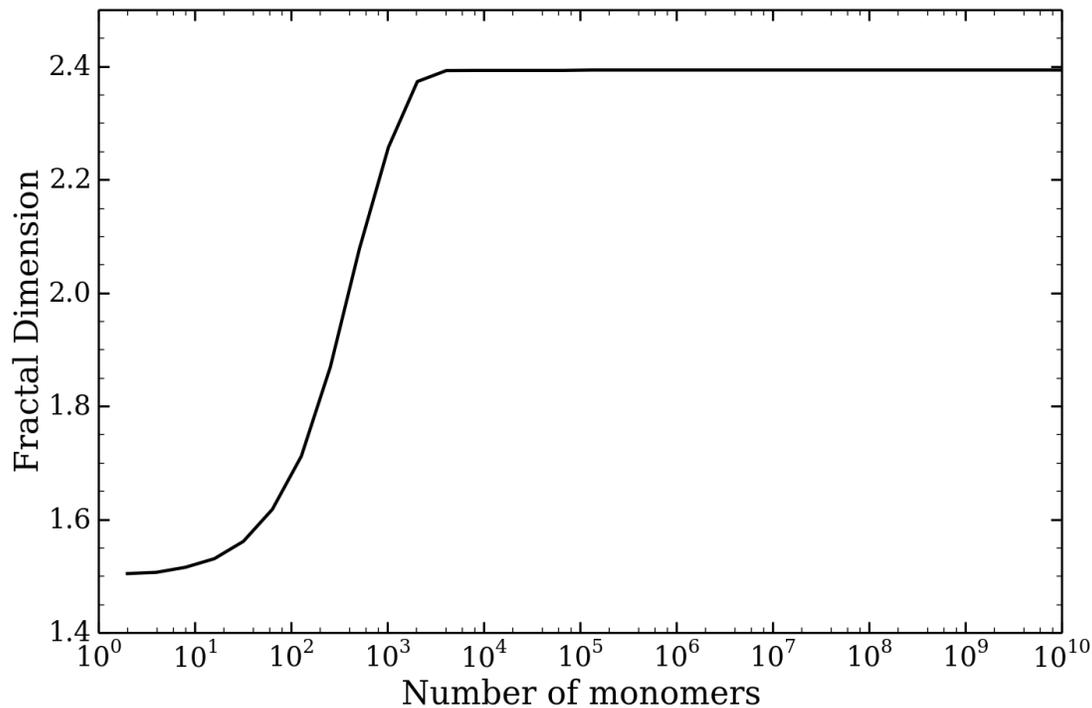

**Figure 1.** Variation of aggregate fractal dimension as a function of the number of monomers in the aggregate.

We consider haze particles composed of hydrocarbon soots, which are likely to survive in hot and warm giant exoplanet atmospheres due to their low volatility at high temperatures, and we use the refractive indices of soot that are presented by Lavvas and Koskinen (2017). As an alternative, we also consider hazes composed of tholins, which have been treated as a proxy for low temperature organic hazes on Titan and Pluto (Khare et al. 1984). As shown in Fig. 2, while the refractive indices of soots are smoothly varying, that of tholins show features at 3 and 6.5 microns, and are lower in value overall. In section 4.2, we compare the effect of tholin and soot hazes on hazy transmission spectra.

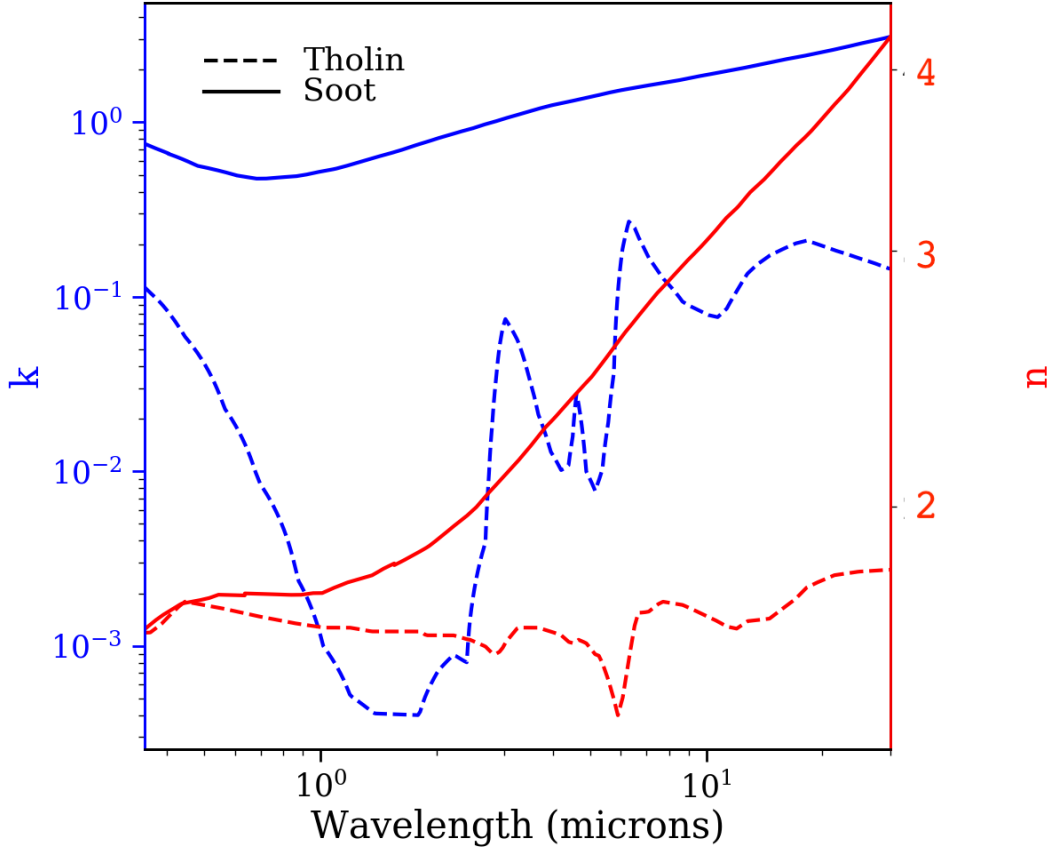

**Figure 2.** Real (n, red) and imaginary (k, blue) refractive indices of tholins (dashed) and soots (solid line).

The optical properties of spherical particles are computed using the Mie scattering code of Grainger et al. (2004), and that of aggregate particles are computed following the methodology of Rannou et al. (1997). The nadir optical depth is defined as:

$$\tau_{nadir,z}(\lambda, z) = \sum_{i=1}^{z} \sigma_{ext}(\lambda, i) n_p, \quad (4)$$

where $\sigma_{ext}$ is the extinction cross section, further defined for spheres as:

$$\sigma_{ext} = Q_e \pi \, r_p^{\,2} \qquad (5)$$

where $Q_e$ is the extinction coefficient. For aggregates, Rannou et al. (1997) treats each monomer as a Mie sphere, and the fractal structure (equation 1) allows for a description of the relative position between monomers in the aggregate. The scattered fields for each monomer are all summed, while taking into account the phase difference of the incident radiation on each spherical monomer as well as the scattered field due to the spacing of monomers. The monomers' relative positions and scattering terms can yield intensities, from which the cross sections are derived. We refer the reader to the Appendix of Rannou et al. (1997) for greater detail regarding this calculation.

We vary the monomer radius ($r_m$), eddy diffusion coefficient ($K_{zz}$), and production rate of haze particles at the top of the atmosphere at 1 microbar to determine their effects on haze opacity. We consider the following evenly log-spaced values: column mass production rate equivalent to $10^8$, $10^9$, $10^{10}$, and $10^{11}$ methane molecules cm$^{-2}$s$^{-1}$ (mass flux of $2.55 \times 10^{-15}$ to $2.55 \times 10^{-12}$ g cm$^{-2}$s$^{-1}$), K$_{zz}$ of $10^8$, $10^9$, and $10^{10}$ cm$^2$s$^{-1}$, and monomer size of 1 and 10 nm. The production rate of photochemical aerosols is highly uncertain, but exoplanets can potentially have higher production rates than those seen on Titan (Horst et al., 2018), which has an estimated haze mass production rate of $0.5 - 2 \times 10^{-14}$ g cm$^{-2}$ s$^{-1}$ (McKay et al. 2001). Our investigated range of values is within those calculated by Lavvas and Koskinen (2017) with a photochemical model. Similarly, the value of K$_{zz}$ is uncertain. Previous studies of exoplanet atmospheres considered values from $10^6$ to $10^{10} cm^2 \ s^{-1}$ (e.g., Kawashima and Ikoma et al., 2018; Lavvas and Koskinen, 2017; Kopparapu et al., 2012, Miguel et al., 2014, Venot et al., 2015, Barman et al., 2015). The range of monomer sizes we consider also spans values used by previous studies. Gladstone et al. (2016) analyzed Rayleigh scattering observations to conclude that monomer sizes of ~10 nm are expected at Pluto. Gao et al. (2017a) considered monomer sizes of 5 and 10 nm in their modeling study of Pluto's haze, and both Lavvas and Koskinen (2017) and Kawashima and Ikoma (2018) used 1 nm monomers for their exoplanet aerosol modeling work. Larger monomers (~50 nm) have been considered for Titan, Jupiter, and additional modeling works (e.g., Tomasko et al. 2008; West & Smith, 1991; Trainer et al. 2006; Wolf & Toon, 2010), as well as produced in laboratory simulations (e.g., He et al., 2018), but we will primarily use the smaller monomer sizes so as to facilitate comparisons to previous exoplanet modeling studies.

We generate temperature-pressure profiles for solar metallicity giant exoplanets orbiting a Sun-like star using a 1D radiative-convective model (e.g., McKay et al., 1989; Marley et al., 1996) that has been previously applied to both hazy and clear atmospheres of solar system bodies, exoplanets, and brown dwarfs (e.g., Marley et al., 1999; Fortney et al., 2008; Saumon & Marley 2008). We vary the distance from the host star (0.05 and 0.20 AU) to obtain two T-P profiles spanning ~400-1750 K, which are shown in Figure 3. The planet's internal temperature is calculated using the results of Thorngren et al. (2018), which uses Bayesian inference and current observations of hot Jupiters to tie internal temperatures to their equilibrium temperatures. Unlike previous studies (e.g., Kawashima and Ikoma, 2018; Lavvas and Koskinen, 2017), our study does not consider photochemistry. Therefore, temperature affects only coagulation and sedimentation rates, as well as the amplitude of spectral features in the transmission spectra due to scale height differences. Photochemical pathways should differ between the 0.05 AU and 0.2 AU cases, as the carbon reservoir at high altitudes for the former is CO while that of the latter is methane (Figure 3). Recent laboratory work from Horst et al. (2018) showed that the rate of photochemical haze production for a high metallicity atmosphere devoid of methane could reach 31% of that of Titan due to the presence of CO. This suggests that CO can replace methane as a progenitor of aerosols, though whether this is applicable to solar metallicity atmospheres is unknown. We investigate the differences in photochemical pathways by considering production rates that vary by several orders of magnitude. An alternative to carbon-based aerosols is the sulfur-based hazes originating from H₂S photochemistry (Zahnle et al., 2016), but they tend to emerge at lower temperatures than those considered here (Gao et al., 2017b).

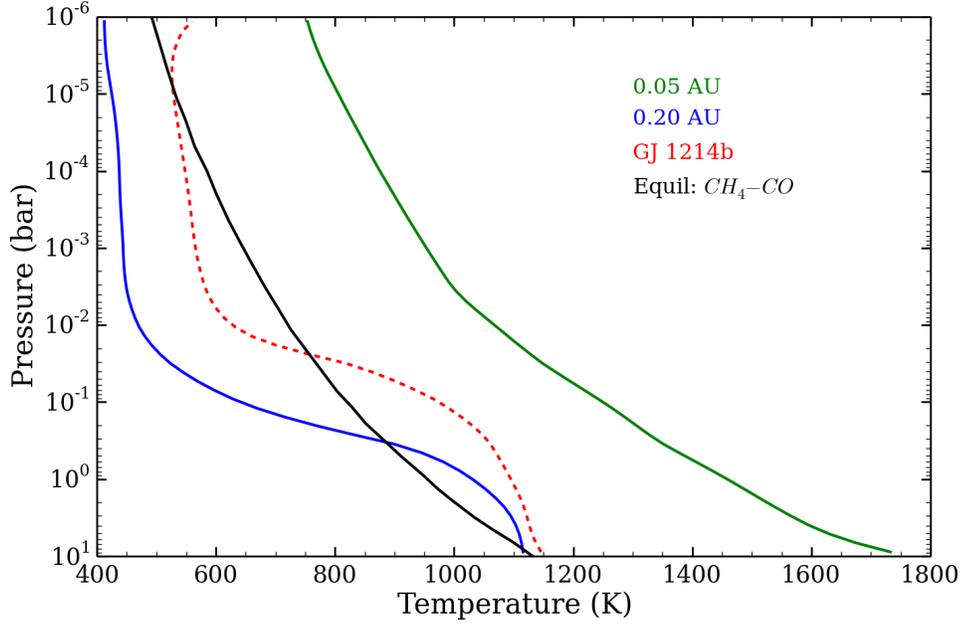

Figure 3. Temperature-pressure profiles of exoplanets at 0.05 AU (green) and 0.20 AU (blue) away from their host star, as well as GJ 1214b (dashed red). The chemical equilibrium [CH₄]=[CO] curve for a solar metallicity atmosphere is shown in black.

We use the same radiative-convective model to generate a TP profile for GJ 1214b, assuming 100x solar metallicity. As in Morley et al. (2013, 2015), we use k-coefficients for a 50x solar metallicity atmosphere multiplied by 2 for the molecular opacity. For the $K_{zz}$ profile, we consider the parameterization specific to the 100 x solar metallicity GJ 1214b case from Charnay et al. (2015), given by:

$$K_{zz} = 3 \times 10^7 P^{-0.4} \quad (6),$$

with P in bars and $K_{zz}$ in cgs units. We consider a column rate of production of 1 nm monomers limited by the diffusion flux of methane into the upper atmosphere,

$$P(z) = n \, K_{zz} \, \frac{df}{dz} \quad (7),$$

with an atmospheric number density $n$ and an eddy diffusivity $Kzz$ at the top of the atmosphere (1 microbar). The methane mixing ratio gradient $\frac{df}{dz}$ is computed assuming total methane photolysis within a scale height (~220 km) above 1 microbar. We calculate the methane mixing ratio at 1 microbar assuming thermochemical equilibrium using the Chemical Equilibrium with Applications code (Gordon and McBride, 1994), yielding 30 ppm. This results in a mass production rate equivalent to $1.3 \times 10^{11} CH_4$ molecules $cm^{-2}s^{-1}$. Note that this value is much smaller than that of Kawashima and Ikoma (2018), who assumed a photon-limited production rate that scaled with lyman alpha flux.

## 3 Results

*3.1 Size Distributions*

Scattering efficiency depends on the particle size relative to the wavelength of light, and therefore the size distributions of haze particles are critical in determining the spectral response. Here, size distributions for aggregate and spherical particles are compared, while varying monomer size, production rate, and the eddy diffusion coefficient. Throughout the atmosphere, aggregate haze particles tend to be larger. This is due to aggregate particles having an irregular shape with a greater cross sectional area than spherical particles of the same mass. This larger cross section allows for an increased number of regions for particles to collide and stick, in comparison to spherical particles. This dependence of coagulation on particle shape is demonstrated by the spatial and size distributions shown in Fig. 4. Aggregate haze particles coagulate to produce a broad distribution centered at a few microns and extending beyond 10 microns, while coagulation of spherical haze particles results in a narrower distribution centered at sizes ~0.3 micron and extending out to only ~1 micron.

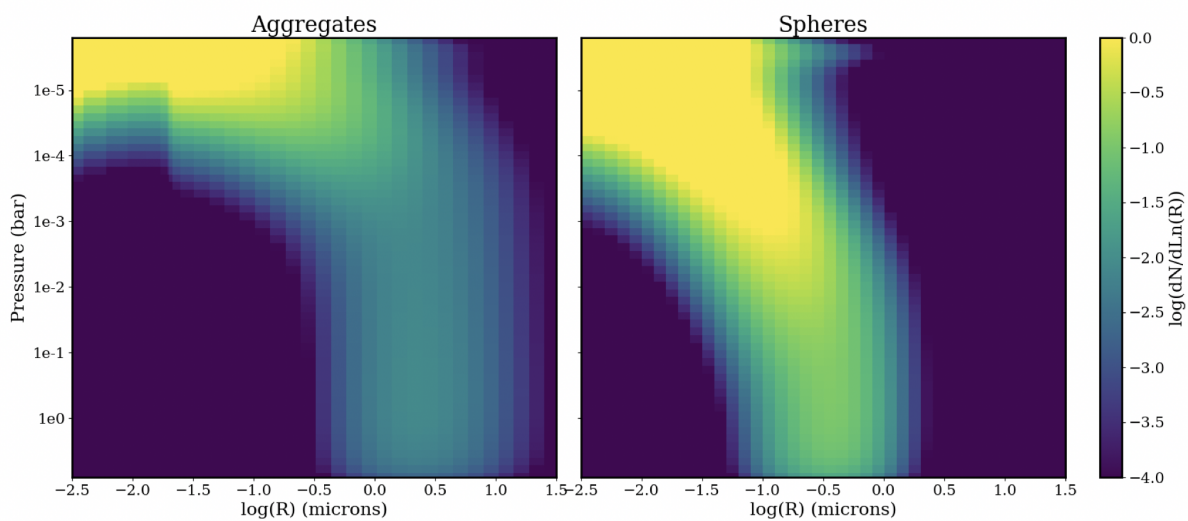

**Figure 4.** Color map of particle size distributions with respect to atmospheric pressure for aggregate particles (left) and spherical particles (right). Parameters were chosen to maximize the haze mass loading, with a mass production rate equivalent to $10^{11}$ methane molecules cm$^{-2}$s$^{-1}$ (maximum), an eddy diffusion coefficient of $10^8$ cm$^2$s$^{-1}$ (minimum), and a monomer size of 1 nm (minimum).

The difference between the two particle types is further demonstrated by comparing the size distributions explicitly in Fig. 5. The aggregate and spherical size distributions peak at ~2.5 and ~0.16 microns, respectively. Due to their greater sizes, it is expected that aggregate particles obscure observations at longer wavelengths more effectively than spherical particles.

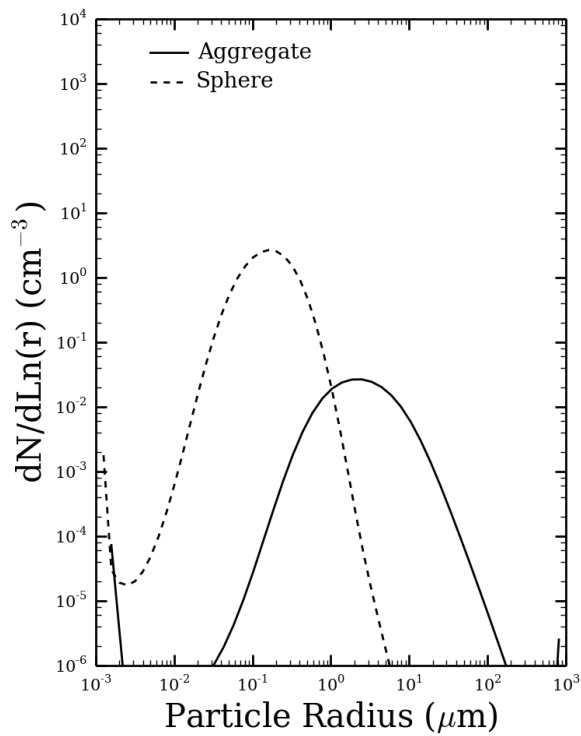

**Figure 5.** Size distributions for spherical (dashed) and aggregate (solid) particles. Parameters were chosen to maximize the haze mass loading, with a mass production rate equivalent to $10^{11}$ methane molecules cm$^{-2}$s$^{-1}$ (maximum), an eddy diffusion coefficient of $10^8$ cm$^2$s$^{-1}$ (minimum), and a monomer size of 1 nm (minimum).

The size distributions are altered by variations in the production rate, eddy diffusion coefficient, and monomer size. An increased production rate allows for more particles to collide and stick together, resulting in the formation of larger particles. As shown by Fig. 6a, the enhanced production rate yields size distributions that peak at radii ~24 and ~8 times larger for aggregate and spherical particles, respectively. Aggregate haze particles respond to changes in production rate more significantly than spherical haze particles. This is due to their larger cross sections and correspondingly increased collision frequency.

Increasing $K_{zz}$ increases the rate of vertical transport of the haze particles, by definition. Allowing the particles to be transported through the atmosphere faster provides decreased time for collisions and coagulation, and this therefore results in smaller particles, as shown in Fig. 6b. At a sufficiently high $K_{zz}$, the haze particles stay small such that no separate coagulation size mode forms.

Monomer size influences the aggregate size distributions due to porosity effects, and the difference in shape of the aggregate distribution is due to the fixed N versus $D_f$ relationship, given by Figure 1. As shown in Fig. 6c, the aggregate size distributions peak at ~2.5 and ~0.77 microns for $r_m$ of 1 and 10 nm, respectively. On the other hand, spherical particles are by definition not porous and their size distributions do not significantly respond to variations in monomer radius. For spherical particles, increasing the monomer size only increases the initial particle size, allowing larger particles to exist at lower pressures, though in lower number densities. Larger particles higher in the atmosphere have different coagulation rates than had they been produced a few scale heights down. This yields slightly offset size distributions for different monomer sizes, as demonstrated in Fig. 6c.

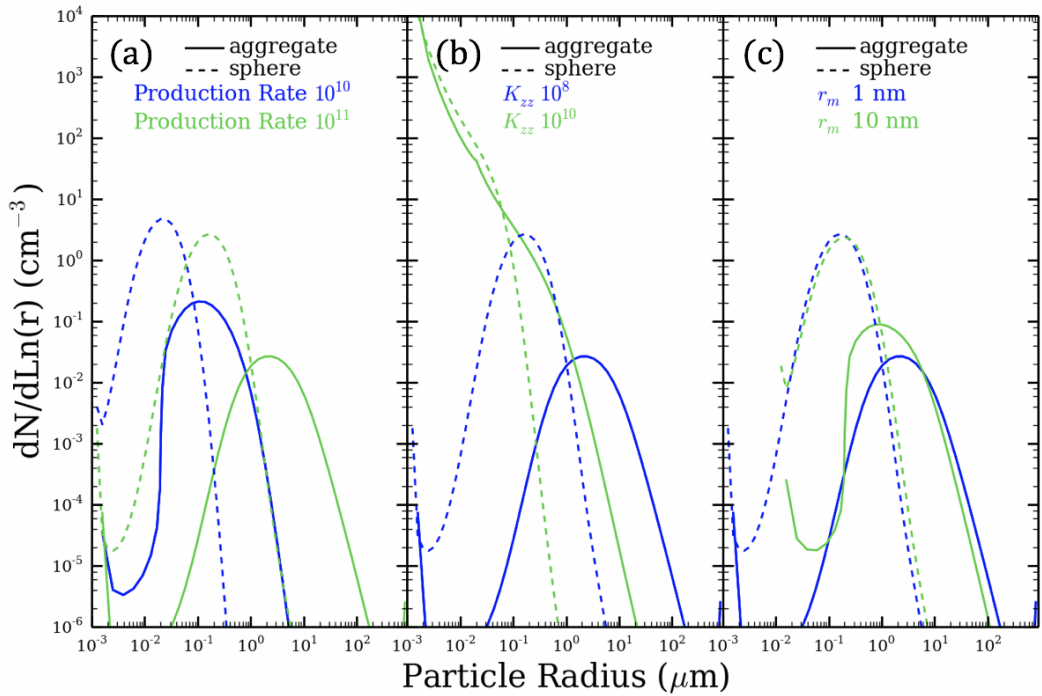

**Figure 6**. Size distributions are compared at ~2.8 mbar (in the middle of our model atmosphere) for spherical and aggregate particles, while varying (a) production rate, (b) diffusion coefficient, and (c) monomer size. Aggregate and spherical haze particles are shown in solid and dashed profiles respectively. Green profiles represent the lower value for the varied parameter and blue represents the greater value. The following parameters were used in each figure: (a) $K_{zz} = 10^8$ cm$^2$s$^{-1}$ and $r_m = 1$ nm, (b) mass production rate equivalent to $10^{11}$ methane molecules cm$^{-2}$ s$^{-1}$ and $r_m = 1$ nm, (c) mass production rate equivalent to $10^{11}$ methane molecules cm$^{-2}$ s$^{-1}$ and $K_{zz} = 10^8$ cm$^2$s$^{-1}$.

### 3.2 Optical Depth

We first calculate the optical depth at each altitude layer. The nadir optical depth at a given pressure level is then calculated by summing the optical depths of the layers above, given by:

$$\tau_{nadir,i}(\lambda) = \sum_{i=1}^{n} \tau_n(\lambda) \qquad (8)$$

Similar to the size distributions, the nadir optical depth is unique to particle shape as well. The wavelength dependence of the spherical haze optical depth is significant, in comparison to that of aggregate hazes, as shown in Fig. **7**. When wavelength is increased from 0.4 to 30 microns, the nadir optical depth for aggregate particles decreases by a factor of ~3 while nadir optical depth for spherical particles decreases by nearly two orders of magnitude, given the same values for the other parameters. Aggregates generally have a greater optical depth than spheres, with the exception of high altitude hazes at a wavelength of 0.4 microns. High in the atmosphere at 10 microbar, spherical particles grow to become moderately opaque at 0.4 microns. By comparison, aggregates grow to larger sizes more quickly, such that, while they can block more light per single particle, their number density is much lower. This allows spheres to be more opaque at 0.4 microns high in the atmosphere. At greater pressures, coagulation of spheres also reduces their number density. Thus, while the extinction coefficient of spherical particles remains small (~2), that of aggregates, computed assuming the cross sectional area of an equivalent-mass sphere, increases (from ~12 to ~20). The change in both density and extinction coefficient allows aggregates to become more opaque deeper in the atmosphere. At 3 microns, however, spheres have a much lower extinction coefficient (~0.2) than aggregates (~32) at the top of the atmosphere, and the larger particles of aggregates are more efficient at scattering long wavelengths of light. Hence, at this wavelength, aggregates are more opaque than spheres at all levels of the atmosphere.

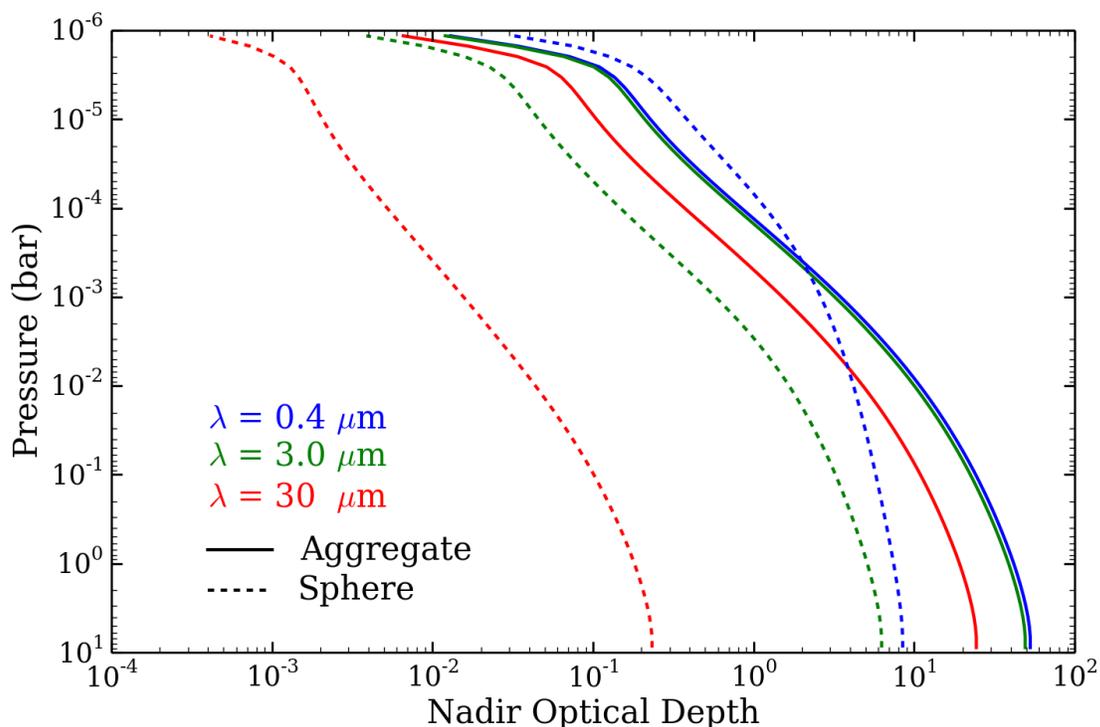

**Figure 7.** Nadir optical depth profiles for aggregate (solid) and spherical (dashed) haze particles at wavelengths of 30 (red), 3 (green), and 0.4 microns (blue). Mass production rate equivalent to $10^{11}$ methane molecules $cm^{-2}s^{-1}$, eddy diffusion coefficient of $10^{8}$ $cm^{2}s^{-1}$, and monomer radius of 1 nm were used.

The pressure level of unit optical depth as a function of wavelength dictates the maximum depth to which observations may probe, particularly for thermal emission; transmission observations probe lower optical depths. As a result of the difference in the wavelength dependence of optical depth of the two particle shapes, aggregate particles are more efficient at obscuring molecular spectral features at long wavelengths than spherical particles (Fig. 8), as aggregate hazes reach unit optical depth higher in the atmosphere and across a broader range of wavelengths. At short wavelengths ($\lambda < 0.5\ \mu m$), spherical and aggregate hazes reach an optical depth of unity at a comparable pressure level in the atmosphere. For example, in comparing opacity at 0.35 and 10 microns in Fig. 8a, unit optical depth occurs ~7 scale heights and <1 scale height deeper in the atmosphere for spherical and aggregate hazes respectively.

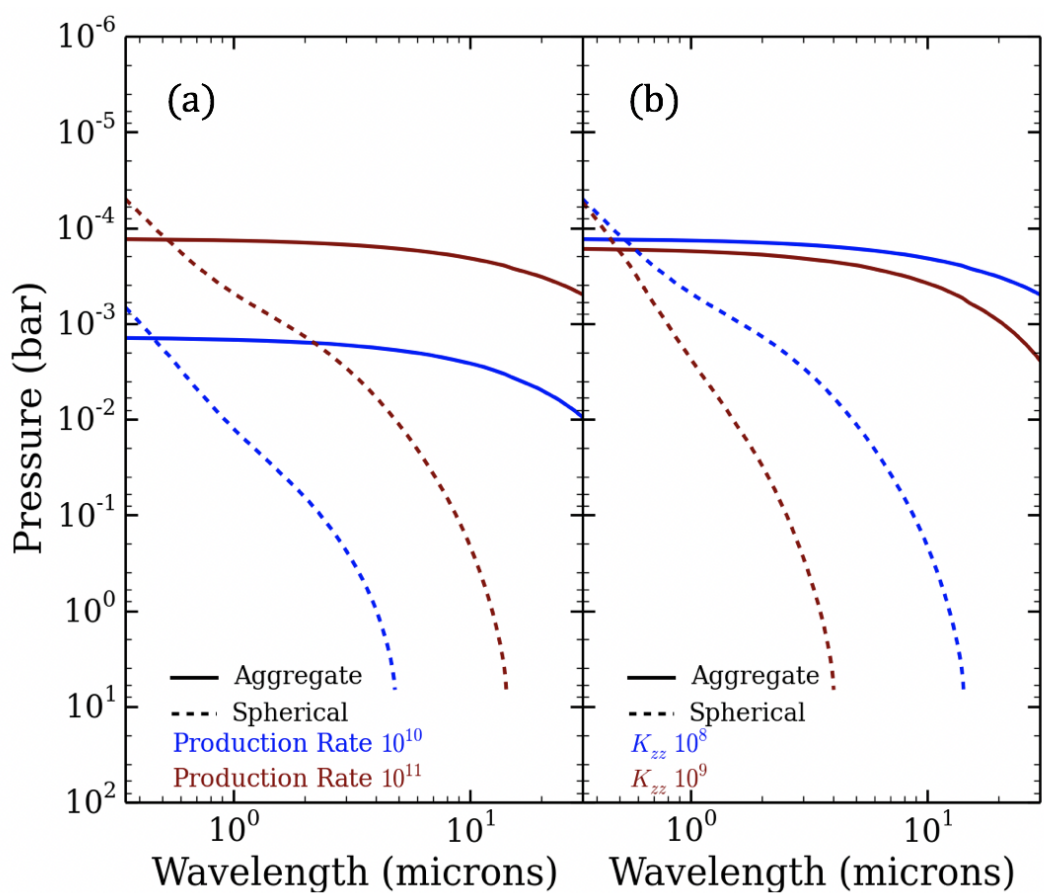

**Figure 8.** Comparison of the pressure levels at which the nadir optical depth reaches unity for aggregate and spherical hazes while varying (a) production rate and (b) eddy diffusion coefficient. The pressure of unit nadir optical depth for aggregate and spherical hazes are depicted by solid and dashed profiles, respectively. Blue profiles correspond to the lower value of the varied parameter, and red profiles correspond to the greater value. The following parameters were used in each figure: (a) $K_{zz} = 10^8$ cm$^2$s$^{-1}$ and $r_m$ = 1 nm and (b) mass production rate equivalent to $10^{11}$ methane molecules cm$^{-2}$ s$^{-1}$ and $r_m$=1 nm.

Increasing production rate increases the number density of large particles at low pressures and results in optical depth reaching unity higher in the atmosphere. The haze particle sizes at the unit optical depth level does not change significantly, however, so the slopes of the unit optical depth level are not affected (Fig. 8a). However, increasing the eddy diffusion coefficient results in smaller particles due to faster transport and decreased time for coagulation, resulting in steeper slopes for the unit optical depth level (Fig. 8b), as smaller particles are less efficient at scattering at longer wavelengths.

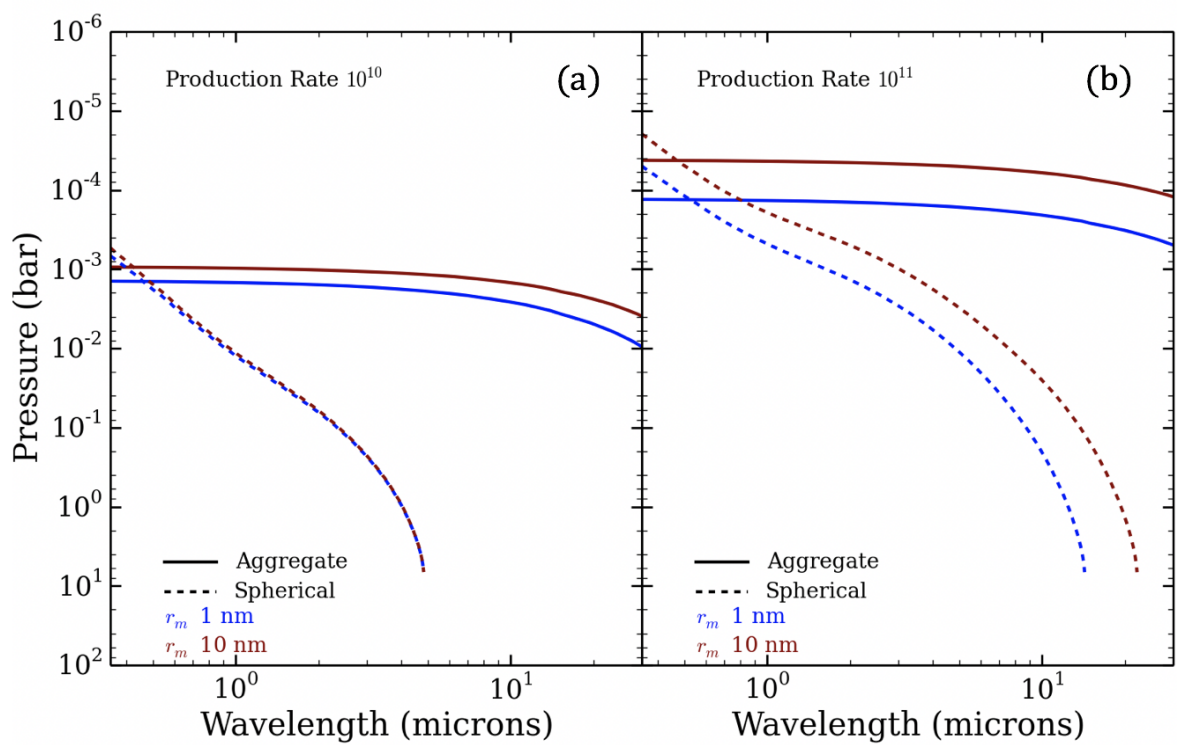

**Figure 9.** Level of unit optical depth with monomer sizes of 1 (blue) and 10 (red) nm for aggregate (solid) and spherical (dashed) haze particles. A production rate of $10^{10}$ methane molecules $cm^{-2} s^{-1}$ and eddy diffusivity of $10^{8} cm^{2} s^{-1}$ were considered.

Increasing the monomer size from 1 to 10 nm leads to a minor increase in the altitude at which unit optical depth occurs in the atmosphere (Fig. 9). For aggregate hazes, the size distribution of haze particles composed of 10 nm monomers peaks at smaller sizes but the distribution is broader. The slope of the unit optical depth level is mainly consistent between the 1 and 10 nm monomer size cases, but unit optical depth occurs deeper in the atmosphere with 1 nm monomers. This effect may be due to the relationship between $N$ and $D_f$, in which for a constant N and particle radius, the fractal dimension increases with an increasing monomer size. However, the vertical shift of this level is relatively small, and the uncertainties regarding the relationship between $N$ and $D_f$ likely dominate any conclusions one could make about a difference in haze opacity caused by monomer size. Also, the effect of monomer size on the level of unit optical depth is found to depend on production rate: given P = $10^{10}$ molecules $cm^{-2} s^{-1}$ (Fig. 9a), a minor shift in haze opacity between the 1 and 10 nm monomer cases is observed for aggregates and no difference is observed for spherical particles; increasing the production rate to $10^{11}$ (Fig 9b) results in a greater shift between the two aggregate cases and causes a shift to develop between the spherical 1 and 10 nm cases. This dependence on production rate exemplifies the effect of the uncertainties regarding N vs $D_f$, and suggests that monomer size alone may not have a significant effect on haze opacity. We also tested the impact of having 50 nm monomers on haze optical depth and found that it continued the trend seen previously with 1 nm and 10 nm monomers, in that the optical depth increased, and so for clarity we do not include those results here.

It is common for haze particles in planetary atmospheres in the solar system to have a potential charge on their surface, or charge density parameter, of up to 30 $e^- \mu m^{-1}$, or 30 electron charges per unit particle radius on a particle's surface (e.g., Borucki et al. 1987; Lavvas et al., 2010; Larson et al. 2015; Gao et al. 2017a). A nonzero charge density parameter can affect coagulation rates and therefore particle size. We therefore considered the effect of varying charge density from 0 to 30 $e^- \mu m^{-1}$. For spherical particles, the presence of charge results in slightly smaller particles and decreased opacity at long wavelengths, while for aggregates the effect was insignificant, as shown in Fig. 10. Despite the larger size of aggregate particles, we surmise that the corresponding coagulation rate is sufficiently high that charge effects become negligible. We consider zero charge density for all other results in this study.

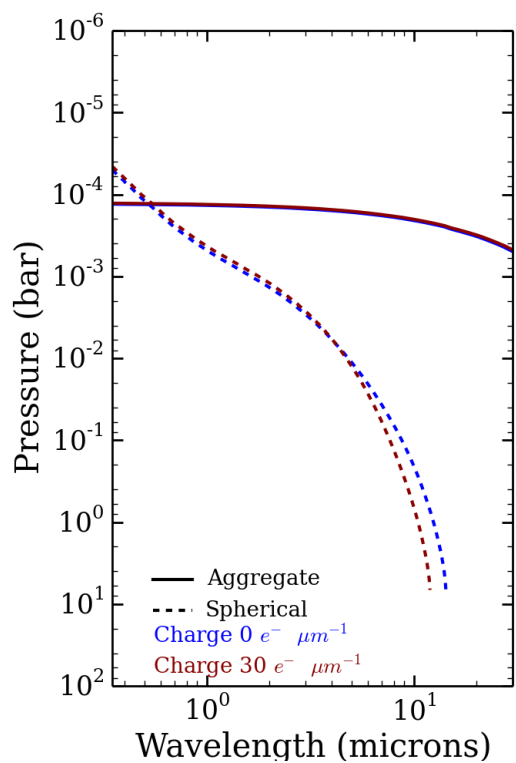

**Figure 10**. Comparison of the pressure level at which unit optical depth occurs when charge is varied from 0 $e^- \mu m^{-1}$(blue) to 30$e^- \mu m^{-1}$(red) for both aggregate (solid) and spherical (dashed) haze particles. The following parameters are used to generate these profiles: mass production rate equivalent to $10^{11}$ methane molecules $cm^{-2}s^{-1}$, eddy diffusion coefficient of $10^8$ $cm^2s^{-1}$, and monomer size of 1 nm.

Several cases produced optically thin hazes that never reached unit nadir optical depth in the atmosphere, such as hazes yielding from mass production rates $\leq 10^9$ methane molecules cm$^{-2}$s$^{-1}$ and $K_{zz} = 10^{10}$ cm$^2$s$^{-1}$. Transits probe slant optical depths and may detect opacity not viewed from the top down. This is shown in the generated transmission spectra in section 4.1.

### 3.3 Application to GJ 1214b

Similar to the results shown in our giant exoplanet cases, size distributions of aggregate hazes in the atmosphere of GJ 1214b extend to larger particle sizes than that of spherical hazes, while a greater production rate leads to larger particles for both aggregate and spherical hazes, as shown in Fig. 11a. Likewise, the depth of unit nadir optical depth is less wavelength dependent for aggregate than spherical hazes, and a greater production rate produces hazes of unit optical depth significantly higher in the atmosphere, as shown in Fig. 11b. From these results, it is clear that aggregate and spherical hazes would have significantly different effects on the transmission spectrum of GJ 1214b.

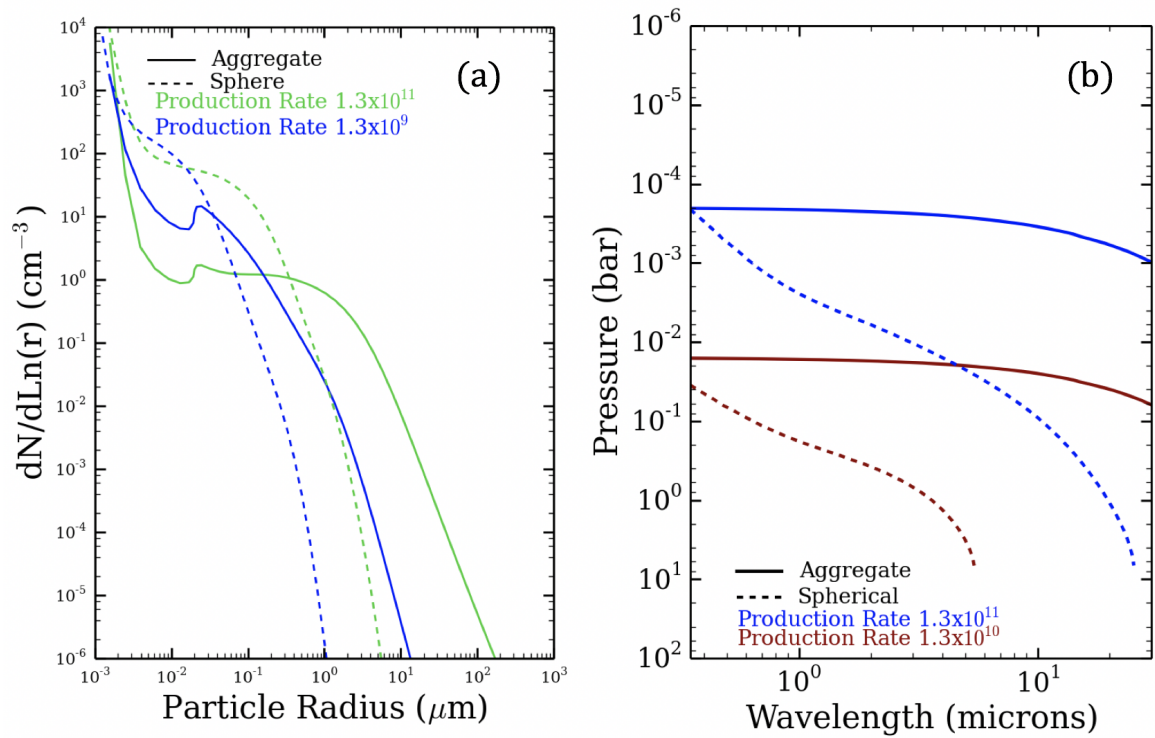

**Figure 11.** (a) Particle size distributions for GJ 1214b with aggregate (solid) and spherical (dashed) hazes. A mass production rate equivalent to $1.3 \times 10^{11}$ (green) and $1.3 \times 10^9$ $CH_4$ molecules $cm^{-2}s^{-1}$ (blue) were considered. (b) The depth of unit nadir optical depth of GJ 1214b for aggregate (solid) and spherical (dashed) hazes. A mass production rate equivalent to $1.3 \times 10^{11}$ (blue) and $1.3 \times 10^{10}$ $CH_4$ molecules $cm^{-2}s^{-1}$(red) were considered.

## 4 Discussion

*4.1 Transmission Spectra of Warm Giant Exoplanets and GJ 1214b*

Since photochemical hazes can dominate spectra of exoplanet atmospheres, we follow the methodology of Fortney et al. (2003, 2010) to generate synthetic transmission spectra to explore the variety of observations possible given the parameter space we have considered in this study. We find that aggregate hazes can generate flat spectra, similar to condensate clouds, since aggregate hazes allow for larger particles to form. By contrast, we are unable to produce flat spectra with spherical haze particles despite the large range in the values considered for production rate, eddy diffusivity, and monomer radius. For example, in Fig. 12a, high production rate ($10^{11}\ cm^{-2}s^{-1}$) and low eddy diffusivity ($K_{zz} = 10^8 cm^{-2}s^{-2}$) results in high opacity hazes for both aggregate and spherical haze particles. The spectra with aggregate haze particles are nearly flat across all wavelengths, while the spectra with spherical haze particles are sloped. In hazy conditions, spherical haze particles can obscure most molecular spectral features at wavelengths shorter than ~1 micron, but with increasing wavelength, spectral features become more distinct.

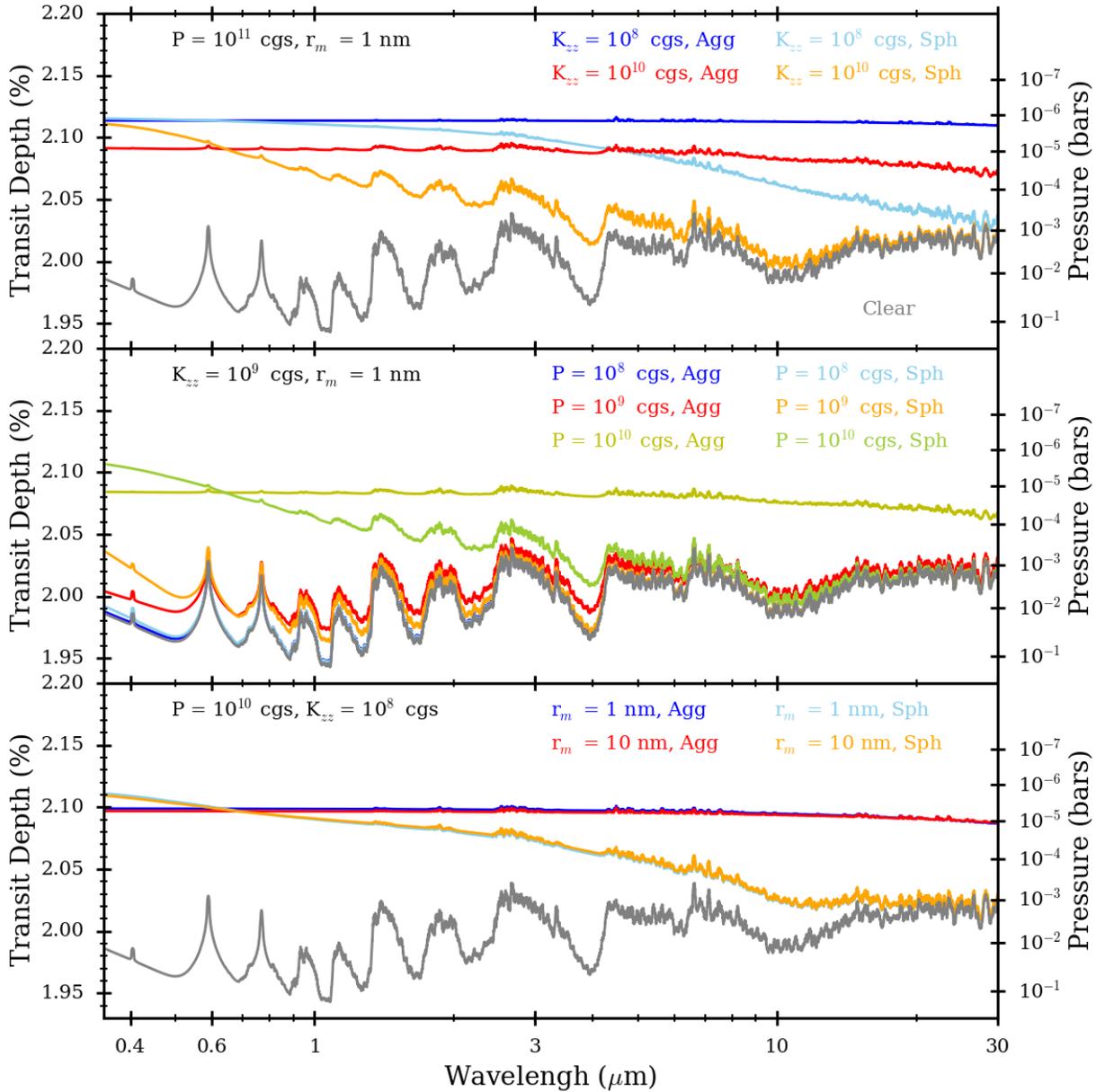

**Figure 12.** Computed transmission spectra for the 0.05 AU atmosphere with hazes considering: (a) $K_{zz}$ of $10^8$ (blue and light blue) and $10^{10}$ $cm^{-1}s^{-1}$ (red and orange) for aggregate (blue and red) and spherical (light blue, orange) hazes. Mass production rate equivalent to $10^{11}$ methane molecules $cm^{-2}s^{-1}$ and $r_m = 1$ nm are used. (b) Mass production rates equivalent to $10^8$ (blue and light blue), $10^9$ (red and orange), and $10^{10}$ (yellow and green) methane molecules $cm^{-2}s^{-1}$ are considered for aggregate (blue, red, and yellow) and spherical (light blue, orange, and green) haze particles. $K_{zz}$=$10^9$ $cm$ $s^{-2}$ and $r_m = 1$ nm are used. (c) Monomer radii of 1 (blue and light blue) and 10 nm (red and orange) are considered for aggregate (blue and red) and spherical (light blue and orange) haze particles. Mass production rate equivalent to $10^{10}$ methane molecules $cm^{-2}s^{-1}$ and $K_{zz}$=$10^8$ $cm$ $s^{-2}$ are used. The transmission spectrum of the clear atmosphere is shown in grey.

The effect of varying eddy diffusivity and production rate on transmission spectra mirrors that on the unit optical depth pressure level. Smaller molecular features are produced with greater production rates and lower eddy diffusivity values due to increased haze opacity. In contrast, varying monomer size results in the opposite effect in transmission as in the nadir view. This is caused by the different regions probed by the two viewing geometries, as well as the more complicated relationship between monomer size and haze opacity, as shown in Fig 13. 1 nm monomer cases show a jump in opacity at the top of the atmosphere before a more gradual increase at higher pressures, while 10 nm monomer cases show a more smooth increase in optical depth. The "jump" for 1 nm monomer cases likely arises due to the forced constant mass flux at the top of the atmosphere, which causes 1 nm monomer haze particles to be more numerous there. Since the coagulation rate is proportional to the square of the number density, 1 nm monomer cases (and the equivalent cases for spherical particles) coagulate faster at the top of the atmosphere, producing larger particles that increase haze opacity. Lower in the atmosphere, farther away from the haze production site, coagulation reaches equilibrium, allowing size distributions and haze opacity to become consistent across monomer radius (as observed for the lower production rate, spherical haze case in Fig. 13). Greater production rates exaggerate the difference between coagulation rates at the top of the atmosphere between the 1 and 10 nm monomer cases.

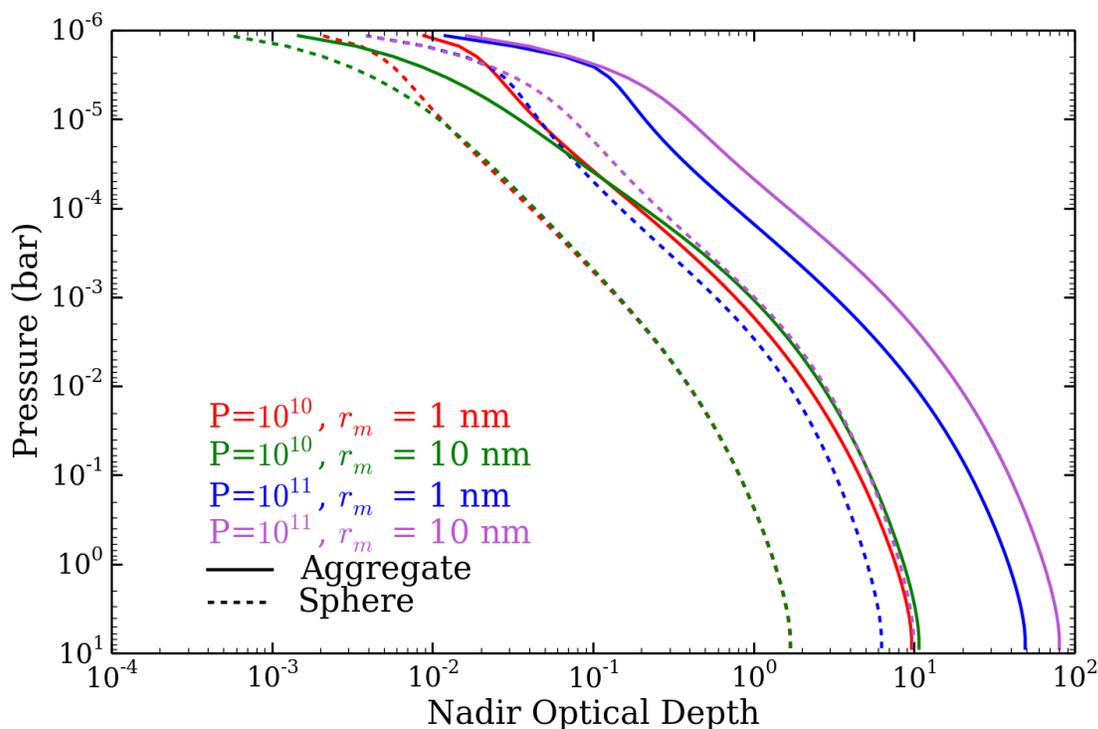

**Figure 13.** Effect of monomer radius on haze opacity depends on production rate and pressure. The nadir optical depth of aggregate (solid) and spherical (dashed) hazes are shown, and production rates of $10^{10}$ (red and green) and $10^{11}$ (blue and purple) methane molecules $cm^{-2} s^{-1}$ and monomer radii of 1 (red and blue) and 10 (green and purple) nm are considered.

It is evident that transmission spectra could be highly sensitive to haze properties. For example, increasing the production rate in Fig. 12b from $10^8$ and $10^9$ $cm^{-2} s^{-1}$ does not significantly change the prominence of molecular spectral features. However, increasing the production rate another order of magnitude, from $10^9$ and $10^{10}$ molecules $cm^{-2}s^{-1}$, results in very hazy conditions, with minimal spectral features. Since the production rate is a poorly constrained parameter for exoplanets (as discussed in Section 1), this sensitivity emphasizes the breadth of observations possible within this parameter space, as discussed by Kawashima and Ikoma (2018).

Spherical and aggregate hazes do not significantly differ across atmospheric temperature profiles with a fixed production rate, eddy diffusivity, and monomer radius. The obscuration of molecular features is comparable for planets at 0.05 AU and 0.20 AU as shown in Fig. 14, with the main difference due to the significantly different scale heights between the two cases. Note that we do not consider photochemistry or atmospheric dynamics in our model, so production rate and eddy diffusivity are not affected. Instead, only the temperature responses of transport and coagulation rates are considered.

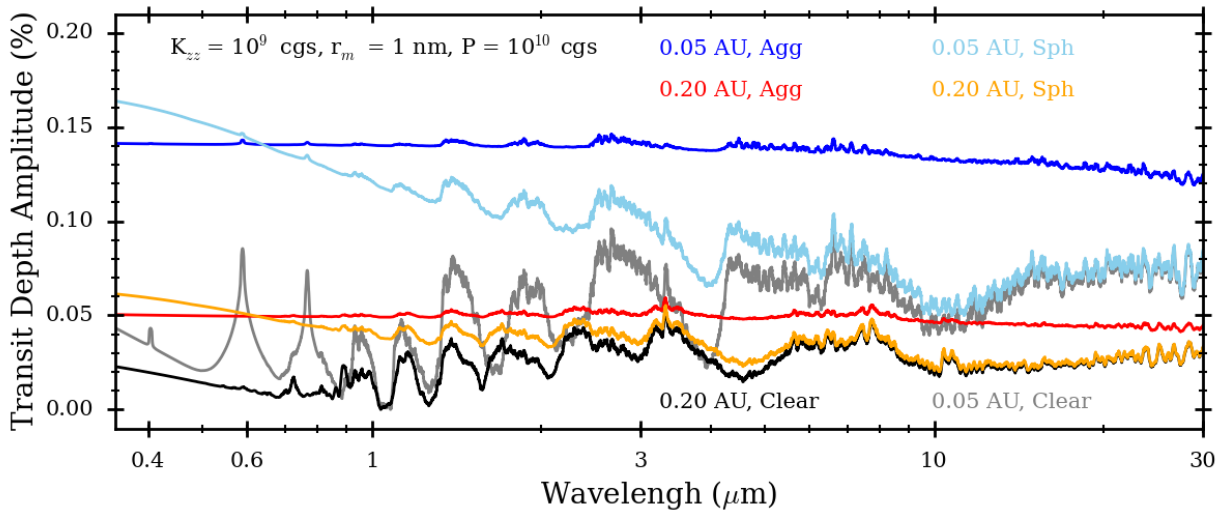

**Figure 14.** Transmission spectra models for planets at 0.05 AU (blue and light blue) and 0.20 AU (red and orange) with aggregate (blue and red) and spherical (light blue and orange) haze particles. The following parameters were considered: $K_{zz} = 10^9 \ cm^2 s^{-1}$, mass production rate equivalent to $10^{10}$ methane molecules $cm^{-2}s^{-1}$, and $r_m = 1$nm. The minimum transit depth value of the clear cases was subtracted from the spectra in order to compare the two atmospheres across different planetary radii.

In order to investigate the effect of aggregate and spherical hazes on the transmission spectra of GJ 1214b, we generate synthetic spectra of the case discussed in section 3.3 with the diffusion limited mass production rate equivalent to $1.3 \times 10^{11}$ methane molecules $cm^{-2}s^{-1}$. Comparing these models to observations from HST WFC3 (Kreidberg et al., 2014) and Spitzer (Gillon et al., 2014) yields a reduced chi squared of 1.58 for the aggregate haze case and 7.29 for the spherical haze case. These results are shown in Fig. 15. For reference, a flat spectrum results in a reduced chi squared of 1.64 (Gao et al. 2018b). The aggregate haze case fits the observations better than a flat spectrum due to $CO_2$ absorption at 4.5 microns in the former case coinciding with an increased transit depth measured in the Spitzer 4.5 micron band. The spherical haze case does not fit the data at all due to the pronounced scattering slope across the entire wavelength range presented. Note that the water feature at 1.4 microns is more pronounced in our aggregate haze model spectra than in the data, suggesting that a slightly higher haze opacity may better fit the data there. This could be achieved with contributions to the haze mass production rate from CO and $N_2$, which are more abundant than $CH_4$ by several orders of magnitude in a 100 x solar metallicity atmosphere. Also, the $K_{zz}$ profile is a model prediction, and slightly lower values could also contribute to higher haze opacity.

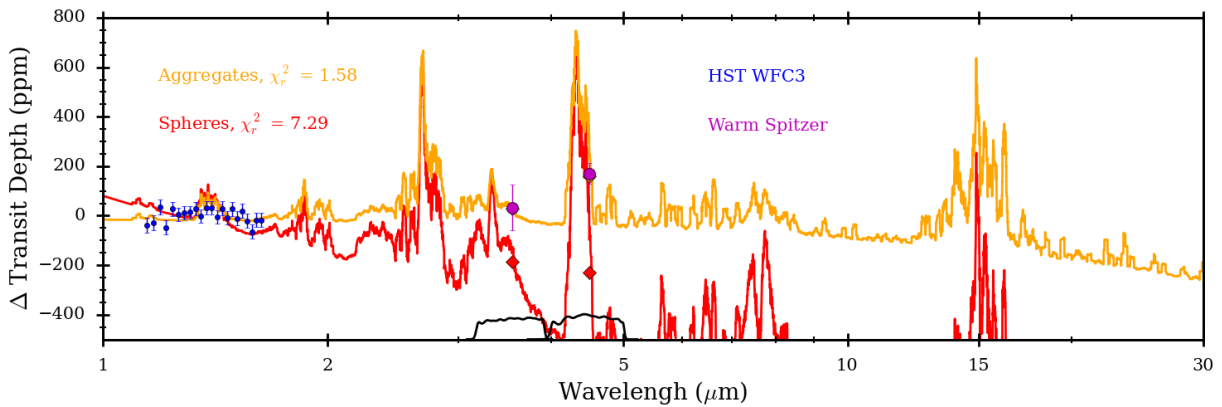

**Figure 15**. Synthetic spectra of GJ 1214b with aggregate (orange) and spherical (red) hazes compared to HST WFC3 and Spitzer observations (blue and magenta; Kreidberg et al., 2014; Gillon et al., 2014). A production rate of $1.3 \times 10^{11} CH_4$ molecules $cm^{-2}s^{-1}$ and 1 nm monomers were considered. The Spitzer filter responses are shown at the bottom of the plot. The Spitzer observations overlay the model values in the Spitzer bands for the aggregate case.

Comparing our GJ 1214b results with previous efforts to fit the observations using microphysical models demonstrates the value in considering aggregate hazes. Kawashima & Ikoma (2018) were able to produce relatively flat transmission spectra using spherical haze particles and reasonable $K_{zz}$ values only when the haze production efficiency was several orders of magnitude greater than that of Titan. Gao et al. (2018b) were able to produce transmission spectra that fit the HST WFC3 and Spitzer points using spherical KCl cloud particles only when the $K_{zz}$ was 1-2 orders of magnitudes greater than those predicted by Charney et al. (2015). Ohno & Okuzumi (2018) considered the $K_{zz}$ profile from Charnay et al. (2015) and spherical KCl cloud particles but were unable to reproduce the cloud top pressures retrieved by Kreidberg et al. (2014) unless the atmosphere was dominated by metals and the cloud particles were porous. However, we note that, even though we are able to reproduce the data with both reasonable $K_{zz}$ values and haze production rates, there exist large uncertainties in haze optical and material properties and photochemical networks that could impact our results.

### 4.2 Tholin Hazes

Photochemical hydrocarbon hazes can have compositions other than soots. For example, hazes at Titan and Pluto have been modeled as tholins (e.g., Gao et al., 2017a). We apply tholin refractive indices to our haze particle distributions and find that their opacity is comparable but less than that of soot hazes, as shown by the synthetic spectra in Fig 16. This is understandable since the imaginary refractive index of tholins is much smaller than that of soots (Fig 2). Unique from soots, tholins have spectral features in the infrared. The most prominent spectral feature occurs at ~6.5 microns and is visible for the spherical case due to the small size of the particles. Tholin hazes are nearly indistinguishable from soot hazes for the aggregate case, as aggregate particles are sufficiently large (e.g. r > 6.5/2$\pi$) and compact in our parameterization of fractal dimension that the spectral feature is muted.

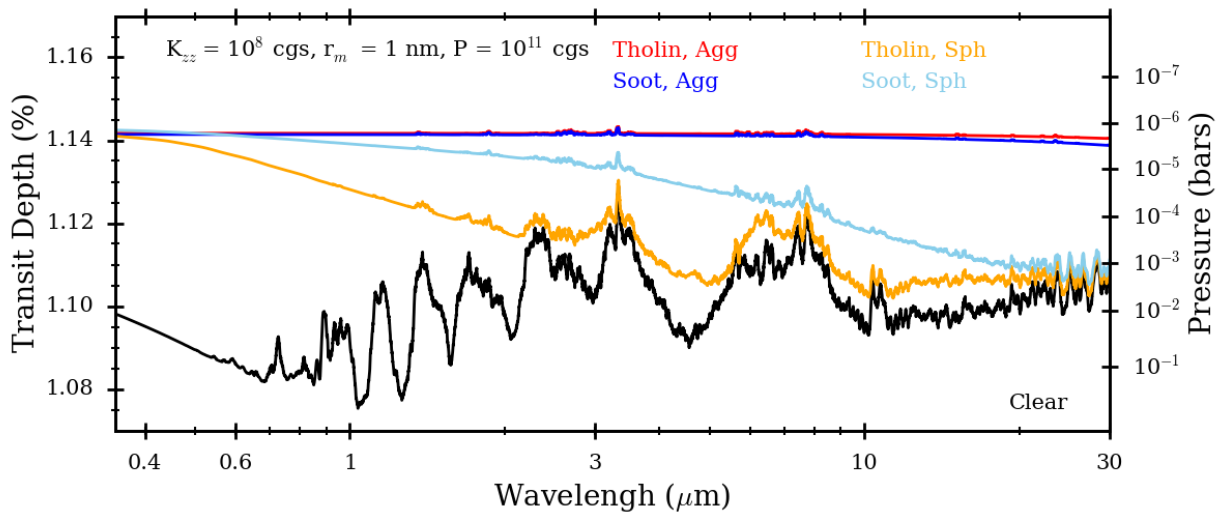

**Figure 16.** Synthetic spectra of the 0.2 AU atmosphere. Clear conditions are shown in black. Tholin aggregate hazes (red), soot aggregate hazes (blue), tholin spherical hazes (yellow), and soot spherical hazes (light blue) are considered. $K_{zz} = 10^8 \ cm^2 s^{-1}$, $r_m = 1$ nm, and P = $10^{11}$ methane molecules $cm^{-2}s^{-1}$ are used to generate these haze cases.

### 4.3 Restructuring of Aggregates in Warm Atmospheres

While aggregate particles are common in solar system atmospheres, it is critical to consider whether they may be restructured or destroyed in the warmer atmospheric conditions considered in this study. We investigate this question by comparing the effective collisional energies of aggregates in our work to the critical energies for aggregate restructuring computed in Dominik & Tielens (1997), which includes the following restructuring mechanisms: sticking of aggregates without visible restructuring; losing monomers upon collision; maximum compression; and catastrophic destruction. A collision is considered to cause "catastrophic destruction" if the colliding aggregates are broken into monomers or very small fragments.

The effective collision energy ($E_{eff}$) is defined as:

$$E_{eff} = \frac{1}{2} M v_{col}^2 \qquad \text{(eq 9)},$$

with an effective mass $M$ given by $M^{-1} = M_1^{-1} + M_2^{-1}$ where $M_1$ and $M_2$ are the masses of the colliding aggregates. We set the colliding velocity to the relative thermal velocity of the aggregates in the atmosphere, as with Lavvas and Koskinen (2017):

$$v_{col} = \sqrt{\frac{8kT}{\pi M}} \qquad \text{(eq 10).}$$

Combining Eqs 7 and 8, we find that $E_{eff}$ is only dependent on temperature. The rolling and breaking critical energies are given by $log\ E_{crit} = A\ log(R)\ + B$ with an effective radius $R$ given by $R^{-1} = R_1^{-1} + R_2^{-1}$ where $R_1$ and $R_2$ are the radii of the monomers. A and B are constants depending on the material, and for this analysis, we consider the values for graphite and polystyrene given by Dominik & Tielens (1997) since these materials are composed of carbon and hydrogen, similar to the haze composition we consider. The critical energies for graphite are defined as:

$$log\ E_{roll,G} = log(R) - 4.65$$

$$log\ E_{break,G} = \frac{4}{3}\ log(R) - 2.40$$

The critical energies for polystyrene are defined as:

$$log\ E_{roll,P} = log(R) - 5.45$$

$$log\ E_{break,P} = \frac{4}{3} log(R) - 3.47$$

| | Critical energy | $r_m = 1$ nm | $r_m = 3$ nm | $r_m = 10$ nm |
|---|---|---|---|---|
| sticking without visible restructuring | $E_{eff} < 5E_{roll}$ | $T_G = 31850\ K$ <br> $T_P = 5050\ \ K$ | $T_G = 99560\ K$ <br> $T_P = 15150\ K$ | $T_G = 318530\ K$ <br> $T_P = 50480\ \ K$ |
| losing monomers upon collision | $E_{eff}$ <br> $> 3n_c E_{break}$ | $T_G = 12520\ K$ <br> $T_P = 1070\ \ K$ | $T_G = 54170\ K$ <br> $T_P = 4610\ \ K$ | $T_G = 269750\ K$ <br> $T_P = 22960\ \ K$ |
| maximum compression | $E_{eff} = 1n_c\ E_{roll}$ | $T_G = 6371\ K$ <br> $T_P = 1010\ K$ | $T_G = 19112\ K$ <br> $T_P = 3029\ K$ | $T_G = 63706\ K$ <br> $T_P = 10997\ K$ |
| catastrophic destruction | $E_{eff}$ <br> $> 10n_c E_{break}$ | $T_G = 41740\ K$ <br> $T_P = 3550\ \ K$ | $T_G = 180580\ K$ <br> $T_P = 15370\ K$ | $T_G = 899160\ K$ <br> $T_P = 76530\ \ K$ |

**Table 1.** Critical temperatures at which graphite ($T_G$) and polystyrene ($T_P$) aggregates composed of 1 nm, 3 nm, or 10 nm monomers begin to undergo their corresponding restructuring mechanism. Critical temperatures are derived from critical energies, assuming $n_c = 1$ (or a single contact point within an aggregate).

By comparing the temperatures corresponding to the critical energies to 1750 K ($T_{max}$), the maximum temperature reached in our model atmospheres (Fig. 3), we can assess whether collisions of aggregates lead to any significant restructuring. Table 1 shows the results of this comparison, and reveals that aggregate particles do not undergo restructuring in the atmospheric temperatures considered. While the critical temperatures for the conditions of "losing monomers upon collision" and "maximum compression" for the $r_m = 1$ nm, polystyrene cases exceed $T_{max}$, this is only for 1 contact ($n_c = 1$), or an aggregate composed of only 2 monomers. For both cases, $T_{max}$ is less than a factor of two greater than the critical temperatures, demonstrating that for aggregates composed of three or more monomers ($n_c \geq 2$), the critical temperature will exceed $T_{max}$. The relationship between this critical temperature and the number of monomers in an aggregate is shown in Fig. 17, and it is clear that for more than 2 monomers, the temperatures considered in our models are cooler than that for maximum compression. Therefore, only small aggregates at very high temperatures may possibly undergo restructuring due to collisions. Also, the critical temperatures are much greater for graphite than for polystyrene, and neither of these compositions are fully representative of the hazes considered here.

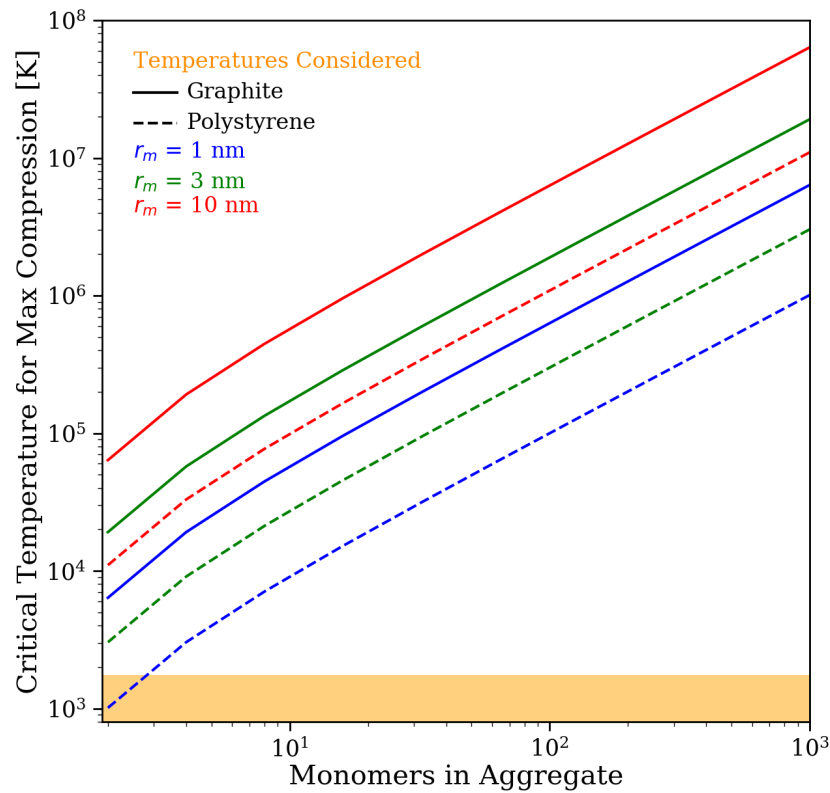

**Fig. 17** Critical temperature for maximum compression as a function of the number of monomers in an aggregate particle. Graphite (solid) and polystyrene (dashed) materials are considered, with monomer radii of 1 nm (blue), 3 nm (green), and 10 nm (red). The range of temperatures considered in our model atmospheres is shown in orange.

While our atmospheric conditions appear to prevent the structures of the aggregates from being significantly affected by collisions, the methodology of Dominik & Tielens (1997) did not consider gas drag. Kataoka et al. (2013) demonstrated that gas drag acting on an aggregate may result in greater compaction than that resulting from collisions. Therefore, we suggest that gas drag may compress larger aggregate particles, allowing for the evolution in $D_f$ that we have assumed in Fig. 1.

*4.4 Haze Production Assumptions*

In this study, we assumed that the haze particles are sufficiently nonvolatile that they do not grow by condensation. This is consistent with the low volatility of soots as indicated by their condensation curves shown in Lavvas & Koskinen (2017). Nucleation is assumed to occur above 1 microbar to form the initial monomers, which is where previous studies have found maximum haze precursor production to occur (e.g., Zahnle et al., 2016; Morley et al., 2013, 2015; Lavvas and Koskinen, 2017; Kawashima and Ikoma, 2018).

Despite our results, hazes composed of aggregate particles do not intrinsically produce flat spectra. By comparing the effective optical depth of aggregate and spherical particles, Wolf and Toon (2010) determined that the spectral behavior of tholin aggregate particles are more sloped compared to equivalent-mass spherical particles in an early earth-type atmosphere. Robinson et al. (2014) found significant wavelength dependence in Titan's haze opacity in transmission, with a slope rising towards shorter wavelengths, even though the haze is composed of tholin aggregates with characteristic particle sizes of 1-2 microns. We provide two possible explanations for differences between these previous studies and our work. First, the different spectral behavior of the imaginary refractive index of tholins are shown in Figure 16 to produce more sloped spectra, at least for spherical particles, compared to the spectra produced by haze with the imaginary refractive index of soots. Second, the particle size distributions of Titan's aggregate haze are different from those of our work. For example, we can compare our size distributions (Figure 4) to those of Lavvas et al. (2010) at pressures probed by transmission spectra. From Robinson et al. (2014) we know transmission spectra of Titan probe altitudes from 100-300 km, while for the P=1e11, $K_{zz}$=1e8, and $r_m$=1 nm case the pressures probed in Figure 12a is ~2e-6 bar. Lavvas et al. (2010) showed through microphysical simulations that the haze particle size distribution at 100-300 km altitude in Titan's atmosphere is bimodal, with a large population centered near ~0.01 microns and a second narrow distribution centered near ~1 micron. In contrast, our particle size distribution shows a tail of slightly larger particles, extending to a few microns, and abundant particles between 0.1-0.5 microns, which are lacking in the results of Lavvas et al. (2010). Our greater number density of larger particles likely contributes to the greater opacity at long wavelengths and the flattened spectra compared to the results of Robinson et al. (2014), and shows that the atmospheres in which hazes develop play an important role in determining their spectral behavior.

We have only considered hazes formed from hydrocarbon chemistry, but Zahnle et al. (2016) showed the significance of sulfur chemistry on the production of photochemical hazes. They found that the free radicals that result from sulfur photochemistry can divert carbon into the stronger bonds of CO and $CO_2$ rather than the weaker bonds of hydrocarbons. While Horst et al. (2018) found that CO can spur haze production under certain conditions, interference from sulfur chemistry may affect the aerosol production rate. Photochemical sulfur hazes can also provide significant opacity to obscure molecular spectral features for planets with effective temperatures < 750 K.

Variations in the optical constants of our considered haze materials would affect our computed haze optical depths and synthetic spectra. For example, Mahjoub et al. (2012) found tholin optical properties to be dependent on the methane concentration of the gas from which the tholins are produced, while Tran et al. (2003) identified minimal dependence on the ratio of methane, ethylene, acetylene, cyanoacetylene, hydrogen, and nitrogen. Instead, hazes produced via photochemistry versus plasma discharge were found to have distinct imaginary refractive indices. Imanaka et al. (2004) found that both the chemical composition and optical properties of tholins produced were dependent on the pressure at which they were deposited. In the context of exoplanets, He et al. (2018) identified that large particle color variations in organic haze material occurred for different temperatures when varying metallicity from 100x to 10000x solar in an exoplanet atmosphere analog. Changes to the imaginary refractive index would affect the slope at shorter wavelengths for the relatively large particles in our work, as shown by the difference between soots and tholins in Figure 16, while differences in the specific spectral response of the material would affect haze spectral features at wavelengths longer than the radius of the haze particles.

Our model ignored photochemistry when varying production rate and $K_{zz}$, but previous studies have found that these parameters do not behave independently. The parent molecules of the hazes (including $CH_4$ and $CO$), as well as haze precursors like more complex hydrocarbon molecules and nitriles are also transported via diffusion. Thus, a high $K_{zz}$ could quickly deplete the upper atmosphere of haze precursors, resulting in decreased haze production. On the other hand, faster upwelling of haze parent molecules could lead to increased haze production. For example, Lavvas and Koskinen (2017) found that increasing eddy diffusivity by an order of magnitude increased mass fluxes of major compounds generated by photolysis at HD 189733b by up to ~3 orders of magnitude, while HD 209458b experienced no change in mass flux. However, Zahnle et al. (2016) found that strong vertical mixing creates a more oxidized environment in the upper atmosphere, resulting in lower production rates of haze precursors. Hence, eddy diffusivity and production rate are not independent parameters, but their dependency is complex.

**5 Conclusions**

Numerous studies have suggested photochemical hazes as an interpretation of exoplanet transmission spectra that show an upward slope towards shorter wavelengths and weak molecular features. Previous works have largely considered hazes composed of spherical particles on exoplanets, while both spherical and aggregate haze particles have been inferred to exist in solar system atmospheres. We used a 1D aerosol microphysics model to investigate the effects of aggregate and spherical haze particles on exoplanet spectra while varying haze production rate, strength of atmospheric mixing, initial haze particle mass, and atmospheric thermal profile. Our results showed that:

- For any given set of parameter values, aggregate haze particles grow to larger sizes than spherical haze particles due to the larger collisional cross section per unit mass of aggregates and the resulting increase in collisional frequency.
- Aggregate haze opacity is gray in the optical and NIR, while spherical haze opacity displays a scattering slope towards shorter wavelengths. Therefore, in high haze opacity cases, aggregate hazes could obscure molecular features in transmission across a wide range of optical and infrared wavelengths, while spherical hazes mostly dominate the optical wavelengths only. We note however that this result is dependent on assumptions of material optical properties and the porosity of the aggregates.
- Increasing haze production rates and decreasing atmospheric mixing rates increased haze opacity at all considered wavelengths due to increased particle number densities, particle radii, or both.
- The effect of monomer radius on haze opacity depends on a myriad of factors including production rate, and changes depending on whether it is viewed in nadir geometry or transmission due to variability in haze opacity with pressure level.

Given reasonable ranges of parameter values, both spherical and aggregate hazes were able to produce a continuum of clear-to-featureless transmission spectra, with spherical hazes generating almost exclusively sloped spectra, and aggregate hazes generating almost exclusively flat spectra, mimicking the effects of condensation clouds. By considering aggregate haze particles, we are able to interpret the flat transmission spectrum of GJ 1214b without the need for extremely high haze production rates or eddy diffusivities, showing the value in taking into account more complex hazes beyond the simple spherical assumption. Our ability to produce differently shaped haze spectra (sloped and flat), as well as the sensitivity of the spectra to uncertain parameters such as production rate suggests the need to better constrain haze properties in exoplanet atmospheres to understand their thermal structure and composition. Additionally, any spectral features from haze particles (e.g., tholins) would be useful in providing size constraints on haze particles; though aggregate particles may be too large and/or compact for spectral features at wavelengths <10 microns to be seen.

## Acknowledgements

We thank P. Rannou for helpful comments regarding aggregate scattering. D. Adams acknowledges the UC Berkeley URAP program. P. Gao acknowledges funding support from the 51 Pegasi b Fellowship in Planetary Astronomy from the Heising-Simons Foundation. This work was performed in part under contract with the Jet Propulsion Laboratory (JPL) funded by NASA through the Sagan Fellowship Program executed by the NASA Exoplanet Science Institute.